\documentclass[aps,prd,groupedaddress,nofootinbib]{revtex4-1}

\usepackage{amsmath,amssymb,bm}
\usepackage{graphicx}
\usepackage{float}
\usepackage{epstopdf}
\usepackage{amsfonts}
\usepackage{amssymb}
\usepackage{amsbsy}
\usepackage{amsmath}
\usepackage{latexsym}
\usepackage{sansmath}
\usepackage{bm}
\usepackage{color}
\usepackage{comment}
\usepackage{csquotes}								
\usepackage{eufrak}
\usepackage{accents}
\usepackage[colorlinks=true,breaklinks]{hyperref}

\newcommand{\eq}[1]{Eq.\,(\ref{#1})}

\def\be{\begin{equation}}
\def\ee{\end{equation}}
\def\bea {\begin{eqnarray}}
\def\eea {\end{eqnarray}}
\def\nn {\nonumber}
\def \p {\partial}

\def \l {\left}
\def \r {\right}
\def \D {\mathcal{D}}
\def \N {\mathcal{N}}

\begin{document}
 
\title{Monte Carlo simulation of cosmologies with dust}

\author{Masooma Ali} \email{masooma.ali@unb.ca} \affiliation{Department of Mathematics and Statistics, University of New Brunswick, Fredericton, NB, Canada E3B 5A3}

\author{Syed Moeez Hassan} \email{shassan@unb.ca} \affiliation{Department of Mathematics and Statistics, University of New Brunswick, Fredericton, NB, Canada E3B 5A3}

\author{Viqar Husain} \email{vhusain@unb.ca} \affiliation{Department of Mathematics and Statistics, University of New Brunswick, Fredericton, NB, Canada E3B 5A3}

\begin{abstract}

The quantum theory of the Friedmann cosmological model with dust and cosmological constant ($\Lambda$)  is not exactly solvable analytically. We apply Path Integral Monte Carlo (PIMC) techniques to study its quantum dynamics using the physical Hamiltonian corresponding to the dust field as a clock. We study (i) quantum fluctuations around  classical paths and (ii) formulate the analogues of the no-boundary  and tunnelling proposals  and simulate the ground state wave functions. For $\Lambda < 0$ a unique ground state wave function exists. For $\Lambda > 0$ the physical Hamiltonian is not bounded below, but the path integral for the propagator is convergent over a range of Euclidean time $T$. We investigate the properties of the convergent propagator. The path integral can be made convergent for all values of $T$ by restricting the integral over paths with action greater than equal to zero. We explore the consequences of such a choice.

\end{abstract}

\maketitle

\section{Introduction}

The problem of quantizing gravity is one at the forefront of theoretical physics. Studies of quantum effects in gravity began soon after the completion of General Relativity (GR), and continues to the present time, spanning almost a century \cite{Rovelli:2000aw}. Despite this, at present  there is no consensus on a theory of quantum gravity, although there are multiple competing ideas. These include Causal Dynamical Triangulations (CDT), Causal Sets, Loop Quantum Gravity (LQG), Asymptotic Safety, Euclidean Quantum Gravity and String Theory (-- see e.g. \cite{Carlip:2017dtj} for an overview of  some of these approaches).

 Within the context of canonical quantization, the path integral approach may be utilized to find solutions to the Wheeler DeWitt (WDW) equation. Specifying solutions to the WDW equation involves choosing appropriate boundary conditions for the equation i.e, prescribing the initial conditions for the universe. In the sum over histories approach this corresponds to the choice of paths (or histories) to be included in the sum. It was demonstrated by Hartle and Hawking that the amplitude for a particular $3$-geometry can be obtained via a semiclassical evaluation of the sum over all $4$-geometries it bounds \cite{Hartle:1983ai}. This is the  ``no boundary proposal."  Several proposals have since been made for both the boundary conditions and the set of paths to be included in the sum to find solutions of the WDW \cite{Vilenkin:1982de, Vilenkin:1986cy, Halliwell:1988ik, Halliwell:1989dy,Feldbrugge:2017kzv}. However, these proposals have all been restricted to semiclassical approximations of the path integral.  
 
Aside from technical difficulties in obtaining a computable theory of quantum gravity, there are conceptual challenges as well. One notable issue is the problem of time \cite{Isham:1992ms}.   Evolution in quantum theory  requires a fixed notion of time, external to the physical system under study. This is  the set of times corresponding to Galilean inertial frames in non-relativistic quantum theory, and the set corresponding to Lorentzian inertial observers in relativistic quantum theory on Minkowski spacetime. In both these cases time is provided by a fixed external kinematical structure, the space(time) metric. In general relativity there are no fixed metrics, and any notion of time must be found within the system. This is the problem of time.

 One way to avoid this issue in canonical quantum gravity is to use an (arbitrary) reference time, whereby a  phase space variable is chosen as a clock, and the evolution of all other phase space variables is with respect to this choice; the Hamiltonian constraint of general relativity is solved classically for the momentum conjugate to the chosen time function to obtain the corresponding (non-vanishing) physical Hamiltonian.  Quantization then proceeds by promoting this physical Hamiltonian to an operator on a suitable Hilbert space. This method, called deparametrization, leads to the reduced phase space of physical degrees of freedom. 

Reduced physical Hamiltonians arising from general relativity are not easy to quantize, except in very special cases. Furthermore, the theories resulting from different  time choices are in general not unitarily equivalent, and different from  Dirac quantization  (see e.g. \cite{Schleich:1990gd}), where the Hamiltonian constraint is imposed as an operator condition.  (There is however at least one example in which the results are equivalent for a particular choice of operator ordering in the Dirac quantization scheme \cite{Maeda:2015fna,Ali:2018vmt}.) 

Despite these issues, there has been recent focus on numerical simulations in order to bypass the difficulties of performing analytical calculations in the various approaches to quantum gravity. Several approaches utilize a sum over histories formulation for the dynamics, each differing in discretization schemes.  Notable examples include CDT \cite{Ambjorn:2010rx}, causal sets  \cite{Rideout:1999ub}, and spin foams \cite{Perez:2012wv}. In addition to making the problem of quantum gravity tractable, numerical simulations within these approaches have provided potentially useful insights into the quantum nature of gravitation \cite{Ambjorn:2005db,Glaser:2014dwa}. 

Here we are interested in exploring the path integral approach numerically within the framework of canonical quantum gravity, with specific application to cosmology, using a matter time gauge.  We propose a method to study the reduced phase space quantization of a closed FLRW universe with non-zero $\Lambda$ and a dust field. This system has no exact solution in quantum theory, except for zero curvature \cite{Husain:2011tm, Ali:2018vmt}. We use the dust field as an internal clock, and  the corresponding  physical Hamiltonian to define a path integral. We then use  PIMC techniques to study the quantum theory. Although the study of a particular mini-superspace model does not amount to the study of a full theory of quantum gravity, it provides a testbed for polishing tools and perhaps inferring some qualitative features of the complete theory. We emphasize that although FRLW quantum cosmology has been studied in several different ways,  the cases we consider have no exact analytic solutions -- all results to date are either semi-classical or other approximations, including in Loop Quantum Cosmology \cite{Agullo:2016tjh}. Therefore the quantum theory of this model with matter is useful to study using PIMC.

Monte-Carlo methods are a staple tool in many areas of physics (e.g. lattice QCD \cite{Gattringer:2010zz} and atomic and nuclear physics). In the context of quantum cosmology, these methods have been applied to cosmological models \cite{Berger1988,PhysRevD.39.2426,Berger:1993fm}. For example Ref.  \cite{Berger:1993fm} contains a study of  ans\"{a}tze for regulating the path integral for mini-superspace models, where the Hamiltonian constraint is not solved classically (although  the lapse ($\N$) is fixed using the condition $\dot{\N} = 0$). Our approach is different in that we deparmetrize first, and then use the resulting physical Hamiltonian for simulations.  Furthermore, in contrast to the pure deSitter models studied in literature \cite{Berger:1993fm,Halliwell:1988ik} our model retains a physical degree of freedom after gauge fixing. 

The outline of the paper is as follows. In Section \ref{model} we define the model and  the associated path integral. In Section \ref{MC-method} we review the PIMC method and detail the Metropolis algorithm as applied to the system under study. In Sections \ref{semiclassical}, \ref{noboundary} and \ref{tunnel}, we provide details of our simulations and results. We conclude in Section \ref{summary} with a  summary of our main results and the prospects for further study of canonical quantum gravity models using similar methods.  (Throughout we work in units with $G=c=\hbar=1$.)

\section{The Model}\label{model}

The action of the theory we study is 
\be
S = \int d^4x \sqrt{-g} (R -2\Lambda) + \int d^4x \sqrt{-g} M\left( g^{ab}\p_a \phi \p_b \phi +1 \right)
\ee
where $\phi$ is the dust field, and $M$ is its energy density. The corresponding canonical action is 
 \bea 
S &=& \int d^3x\,dt [ \pi^{ab} \dot{q}_{ab} +  p_\phi \dot{\phi}  \nonumber\\
&& - \N (\mathcal{H}_G + \mathcal{H}_D ) - N^a (\mathcal{C}_G + \mathcal{C}_D) ], 
\eea
where 
\bea
\mathcal{H}_G &=& \dfrac{1}{\sqrt{q}} \l( \pi_{ab} \pi^{ab} - \frac{1}{2} \pi^2 \r) + \sqrt{q} ( \Lambda-R), \nn \\
\mathcal{H}_D &=& \frac{p_\phi^2}{2 M a^3} + \frac{Ma^3}{2} \l(1 + q^{ab} \p_a \phi \p_b \phi \r), \nn \\
\mathcal{C}_G &=& -D_b \pi^b_a, \nn \\
\mathcal{C}_D  &=& -p_\phi \p_a \phi. 
\eea
Here  ($q_{ab}, \pi^{ab}$) and ($\phi,p_\phi$)  are the gravitational and dust degrees of freedom respectively, and $\N$ is the lapse and $N^a$ is the shift.

We proceed by using the dust field to fix the time gauge \cite{Husain:2011tk, Husain:2011tm}. We set $ t = \varepsilon \phi$ with $\varepsilon^2 = 1$ and solve the Hamiltonian constraint ($\mathcal{H}_G + \mathcal{H}_D \approx 0$)  for the conjugate momentum to the dust field. The physical Hamiltonian is then given by
\be
\label{hamiltonian}
H_p = -\varepsilon \int d^3x \, p_\phi  = \varepsilon \int d^3x\, \mathcal{H}_G.
\ee
Requiring that the gauge be preserved in time fixes the lapse to $\N = \varepsilon $. For now we leave the sign of the lapse undetermined, but we will fix it later; the sign of the lapse determines how the dust energy density ($M$) relates to the physical Hamiltonian: negative lapse corresponds to a positive dust energy density when $H_p >0$   \cite{Ali:2015ftw,Hassan:2017cje}.   

\begin{figure*}
\includegraphics[width = 5in]{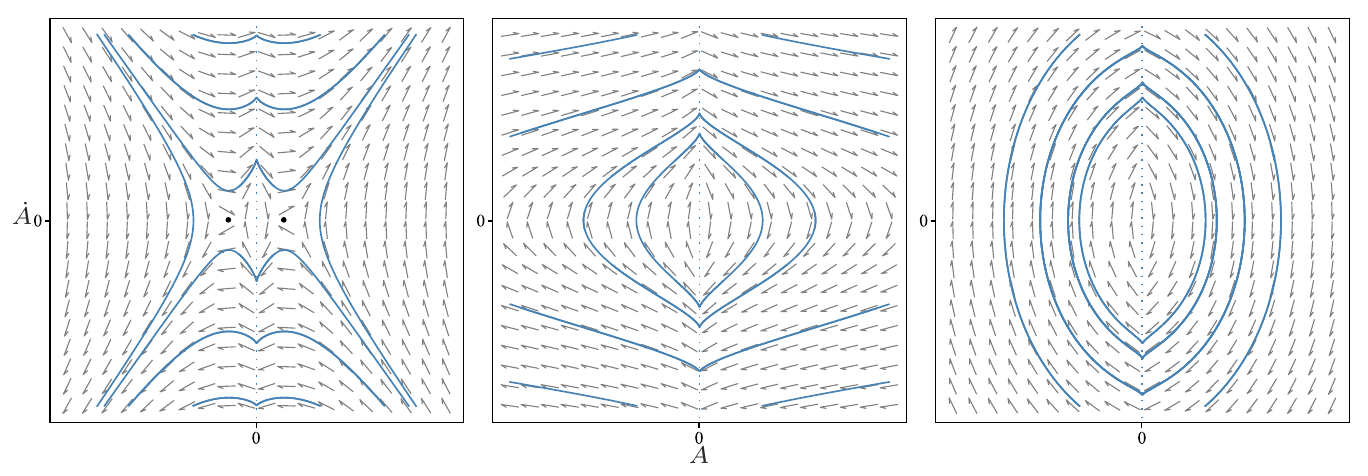}
\caption{Phase portraits for \eq{EOM} with $k= 1$, and $\Lambda = 1$ (left), $\Lambda = 0$ (centre), $\Lambda = -1$ (right). The dotted line at  $A=0$ is the singularity. The black dots in the first frame ($\Lambda >0$) indicate the two saddle points ($\dot{A}=0, dV/dA=0$). }
\label{lorentzian-phase-portraits}
\end{figure*}

This paper is concerned with the homogeneous isotropic cosmological metrics 
\be
ds^2 = -dt^2 + \frac{a(t)^2}{1+ kr^2/4}\left(dr^2 + r^2 d\Omega^2 \right), 
\ee
where $k$ is the spatial curvature. In these coordinates, where the spatial metric is conformally flat,  the canonical theory in the dust time gauge 
 takes a simple form. Let us define $f(r) \equiv 1+kr^2/4$,  $e_{ab} = \text{diag}(1,1,1)$, $h_{ab} = f(r) e_{ab}$, and set 
\bea
q_{ab} &=& \frac{3}{8}A^{4/3}(t) h_{ab}, \nn\\
\pi^{ab} &=& 2A^{-1/3}(t) p_A(t)\sqrt{h}h^{ab},
\eea
where we take the reduced phase space $(A,p_A)$ to be $\mathbb{R}^2$. The various factors ensure that $\pi^{ab}\dot{q}_{ab}\rightarrow p_A\dot{A}$.  

This is an unconventional canonical parametrization where 
$a(t) \sim A^{2/3}(t)$. It has the advantage that the gravitational kinetic term is proportional to $p_A^2$, and the physical Hamiltonian in the dust time gauge takes the form
\be
H_p = \varepsilon \l(-\frac{p_A^2}{2} + \frac{\Lambda}{2} A^2 - k A^{2/3}\r),
\ee
(after the rescalings $\Lambda \rightarrow  \frac{3}{4} \Lambda$, $k \rightarrow  \l(\frac{3}{8} \r)^{\frac{1}{3}} k$). The corresponding action is 
\be
\label{oscillator-action}
S = \int dt \, \varepsilon \l( -\frac{\dot{A}^2}{2} - \frac{\Lambda}{2}  A^2 +  k A^{2/3} \r).
\ee
For $\varepsilon = -1$, \eq{oscillator-action} becomes the action of a particle moving in the potential 
\be
V(A) = -\frac{\Lambda}{2} A^2 + k A^{2/3}.\label{potential}
\ee
   This is the choice we make. The $k=0$  case is then a simple harmonic oscillator for $\Lambda < 0$, and an inverted oscillator for $\Lambda >0$ \cite{Ali:2018vmt}. For non-zero $k$ the equation of motion is
\be
\label{EOM}
\ddot{A}  -  \Lambda A + \frac{2}{3} k A^{-1/3} = 0.
\ee
This is  singular at $A=0$. Fig. \ref{lorentzian-phase-portraits} displays the phase portraits for typical values of $\Lambda$.  The case  $\Lambda > 0$ exhibits  two saddle points where $V'(A)=\dot{A}=0$; near these points the trajectory flow lines on either side are toward or away from the singularity at $A=0$.


We are interested in computing the path integral  
\small
\be
\label{lorentzian-path-integral}
G(A_f,A_i) = \int \D A\, \exp \l( iS[A,\dot{A};k,\Lambda, T] \r),
\ee
 for the action (\ref{oscillator-action}), with $A_i =A(0)$ and $A_f=A(T).$   As for all such integrals, this is oscillatory and so difficult to evaluate. 
 We therefore do a Wick rotation in the dust time gauge:  $t \rightarrow  -it$.  This converts  \eq{lorentzian-path-integral} to 
\be
\label{euclidean-path-integral}
G(A_f,A_i) = \int \D A\, \exp \l(- S_E \r ),
\ee
 where 
\be
\label{euclidean-action}
S_E = \int_0^T dt \l(\frac{\dot{A}^2}{2} - \frac{\Lambda}{2} A^2 + k A^{2/3} \r).
\ee
This  integral  is not analytically tractable for $\Lambda, k \ne 0$.  We therefore proceed by numerically computing the path integral using a Monte Carlo  method. In the following section we define this process in detail.

\section{Monte Carlo method} \label{MC-method}

 We propose to evaluate  the integral (\ref{euclidean-path-integral})  using the Path Integral Monte Carlo (PIMC) technique. The central idea is to generate representative sets of paths that are then weighed with action to calculate the integral. 

As is well known, in classical theory there is a unique on shell-path once the initial conditions are specified, whereas in quantum theory an infinite number of paths  contribute to the Feynman path integral, each with a phase $\exp(iS)$. After Wick rotation, the amplitude $\exp(-S_E)$ may be treated as a probability distribution on the space of paths. The PIMC technique generates a Markov chain of paths from an initial seed path, such that the stationary distribution for the Markov chain is given by the amplitude $\exp(-S_E)$. In order to probe the space of paths effectively we use the Metropolis algorithm for importance sampling. In this approach paths with large positive $S_E$ are  suppressed.

We start by discretizing the action (\ref{euclidean-action}). The time interval from $0$ to $T$ is divided into $N$ steps; $A(t) \rightarrow A_i, i=1..N$. For the time-derivative, we use a forward step, $\dot{A}(t) \rightarrow ( A_{i+1}-A_i )/ \epsilon$. The corresponding discrete action is
\be
\label{discrete-action}
S_E = \sum_{i=1}^{N-1} \epsilon \Bigg[  \frac{(A_{i+1}-A_i)^2}{2 \epsilon^2} - \Lambda A_i^2 + k A_i^{2/3} \Bigg].
\ee
With this discretization, the MCMC method we use proceeds as follows. After fixing an initial path $A^{start}_{\{i\}}$,  (which could be selected by a deterministic or  random rule),  

\begin{enumerate}
 
\item Change a random element of the array: $A_i \rightarrow A^{new}_i = A_i + \delta$, where $\delta \in [-\Delta,\Delta]$ is a random number chosen from a  uniform distribution, with $\Delta$ a fixed parameter;

\item Calculate the change in the Euclidean action: $\Delta S = S_{new} - S$;

\item Accept or reject this change. If $\Delta S  \leq 0$,  the change is accepted, otherwise  it is accepted with a probability $\exp(-\Delta S)$; if  accepted,  the selected element is updated: $A_i := A^{new}_i$;

\item Repeat $n$ times the steps $1-3$. This defines one Monte Carlo (MC) iteration.

\end{enumerate} 

\begin{figure}
\includegraphics[width = 4in]{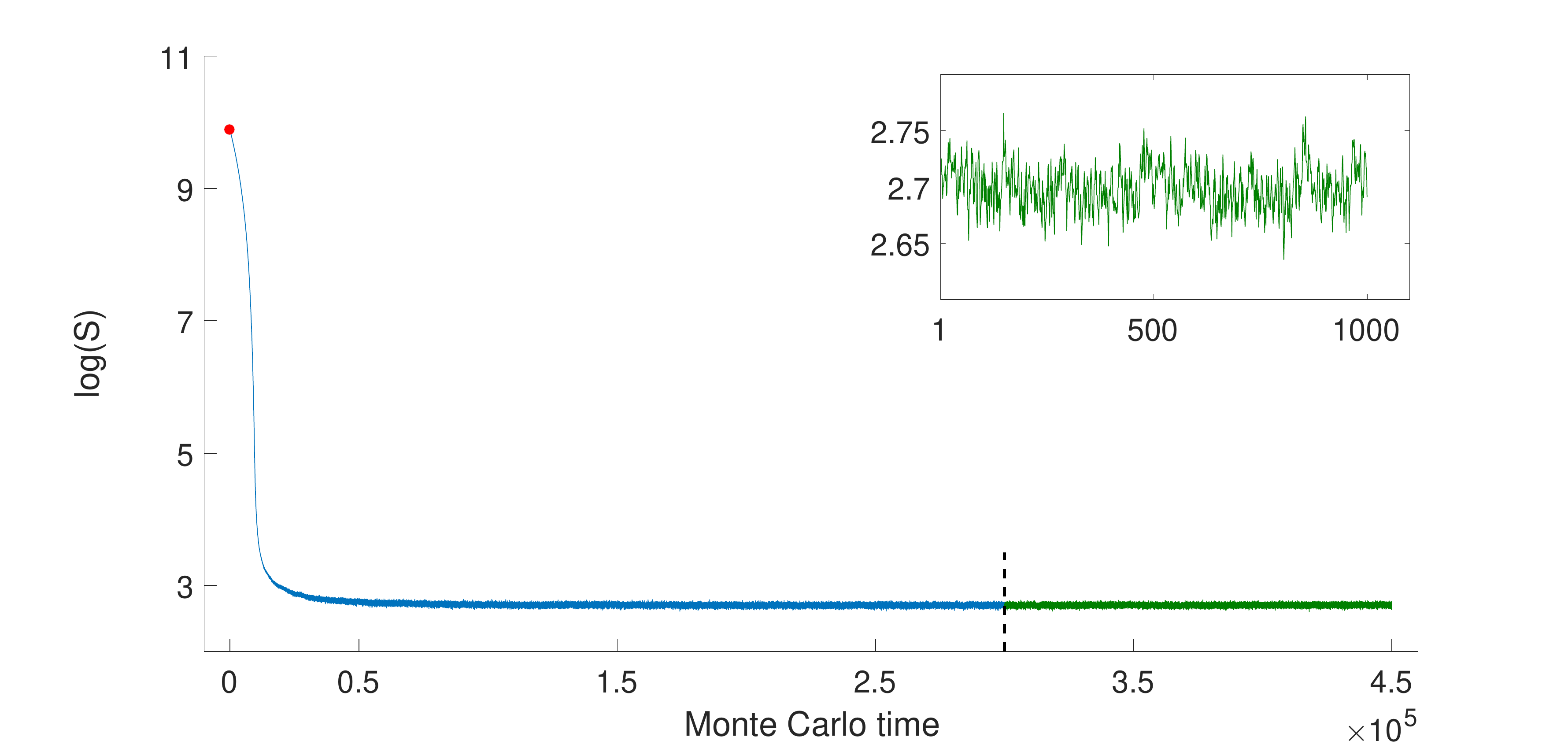}
\caption{Log of the action $S$ vs Monte Carlo time, from one representative run. The red dot shows the starting value of the action (for a random initial path), the blue curve shows the action during thermalization, the dashed black line marks the point where measurements are started, and the green curve shows the action values during measurement. It is clear that the action achieves thermalization after around 100,000 thermalization steps. In the inset, the last thousand samples taken during measurement are plotted, to show the variations in the action around its mean value.}
\label{sample-action}
\end{figure}

An important element in each run is thermalization. This consists of performing a number $N_{therm}$  of MC steps until the action is thermalized. This is to ensure that the MCMC process has lost memory of the starting point before sample path selection begins. $N_{therm}$ is chosen so that the order parameter used (the Euclidean action in our case) reaches an equilibrium value, up to small fluctuations.  Fig. \ref{sample-action} shows the thermalization of the action for a representative run for a given set of MC parameters, with various $A^{start}_{\{i\}}$. It is evident that the action converges to approximately the same value.  

Once thermalization is complete, a number  $N_{MC}$  of MC steps is carried out. Paths are selected as part of the sample every $N_{skip}$ steps, a number determined so as to reduce autocorrelations. After a sufficient number of sample paths are gathered through this process, computation of   propagators and expectation values of observables can proceed. 

At this stage, it is important to note a key difference between the cosmology we are studying and conventional quantum mechanical systems: The potential  (\ref{potential}) is not bounded below when either $\Lambda > 0$, or $k <0$, or both. Thus the Hamiltonian is not bounded below and the Euclidean path integral does not appear to converge. We consider two methods to deal with this issue:

{\bf 1.} Convergence of path integrals of systems with Hamiltonians unbounded below were investigated
in \cite{Carreau:1990is}, with the conclusion that for potentials diverging at most as  $-\omega^2 x^2$, the path integral is convergent if the Euclidean time $T$ is less than $\pi/\omega$. The argument is worth summarizing as it applies directly to our system. Let $S_0$ denote the kinetic term   in \eq{euclidean-path-integral}. Then 
\bea
G(A_f,A_i,T) &=& \int \D A\  e^{-S_E} =  \int \D A\, e^{-S_0} e^{ \int_0^T dt \l( \Lambda A^2/2 - k A^{2/3} \r)} \nn \\
&\leq & \int \D A\, e^{-S_0} e^{(\Lambda T/2) \text{max}[A^2]}
\eea
where max$[A^2(t)]$ denotes the maximum value of $A^2(t)$ along each path $A(t)$ between $A_i=A(0)$ and $A_f=A(T)$;  the last inequality follows since including the $A^{2/3}$ term reduces the integral. To further constrain the r.h.s., we must bound the expectation value 
$\langle e^{(\Lambda T/2) \text{max}[A^2]} \rangle$ with the weight $e^{-S_0}$  to include all paths between $A_i$ and $A_f$ 
(in time $T$) that exceed the value $\zeta \geq \text{max}(A_i,A_f)$. This requires  an integration over $\zeta$ from $\text{max}(A_i,A_f)$ to $\infty$. This is accomplished by noting the following: (i) The amplitude from $A_i$ to $A_f$ in time $T$ with respect to the free action 
$e^{-S_0}$ is  $
 W_F(A_i,A_f,T) = \int \D A\  e^{-S_0} =  \exp\left\{-(A_f-A_i)^2/2T \right\}/\sqrt{2\pi T}
$, (ii) by the reflection principle, for paths crossing $\zeta$, $W_F(A_i,A_f,T)  = W_F(A_i, 2\zeta- A_f,T)$, (iii) the amplitude for all paths 
that lie between $\zeta$ and $\zeta + d\zeta$ is therefore $W_F(A_i, 2\zeta- A_f,T) - $ $W_F(A_i, 2(\zeta+d\zeta)- A_f,T)$ $\approx$
$(d/d\zeta) W_F(A_i, 2\zeta- A_f,T)$. Using this last expression to compute $\langle e^{(\Lambda T/2) \text{max}[A^2]} \rangle$, we arrive at the bound 
\be
G(A_f,A_i,T) \le \int_{\text{max}(A_i,A_f)}^\infty d\zeta\ (2\zeta - A_i-A_f) \sqrt{2/\pi T^2} \ e^{-(A_i+A_f - 2\zeta)^2/2T}  e^{\Lambda A^2T/2} 
\ee
The integral in the last inequality is convergent if $T < \pi/\sqrt{\Lambda}$, and thus $G(A_f,A_i,T)$ is bounded for all 
$T < \pi/\sqrt{\Lambda}$.  

We use this result to calculate the amplitude for the universe to expand from nearly zero volume to some finite volume in section \ref{semiclassical}. In sections \ref{noboundary} and \ref{tunnel}, though we do not fix the end point $A_f=A(T)$ we find the integral is convergent for some values of $T$. We calculate the no boundary wave function and the tunnelling wave function for these $T$ values.

{\bf 2.} An alternative way to proceed is by restricting the integral to the set of paths on which the Euclidean action is manifestly positive or zero. 
(This is similar to the approach in  \cite{Berger:1993fm} where the vacuum model was studied without solving the Hamiltonian constraint classically). This is motivated by the duality between Euclidean quantum field theory and statistical mechanics (see e.g. \cite{McCoy:1994zi}):  The Euclidean action of the quantum theory is akin to the Hamiltonian of a stat-mech system with time identified as (inverse) temperature ($\beta$),
\be
\int \D A ~~ e^{-S_E} \sim \sum e^{-\beta H}.
\ee
Requiring that this Hamiltonian (our Euclidean action) be positive,
\be
S_E \geq 0
\ee
fixes the ground state at $S_E = 0$ and yields a convergent integral.

A necessary consequence of this regularization is that for $\Lambda >0$ (or $k < 0$) there is no unique ground state. In fact an infinite number of degenerate vacua exist since there is an uncountably infinite number of paths which yield $S_E = 0$. Thus the physical results depend quantitatively on whichever subset of vacua the MCMC process converges to in each run. However despite this ambiguity, there are qualitative features common to different vacua which we explore in the following sections; we can compare the averages of observables calculated over different subsets of vacua to extract common features\footnote{This situation with infinite degenerate vacua can be compared to another physical system: A particle in a Mexican hat potential: $V(x,y) = (x^2 + y^2 - a^2)^2$, where there is also a one-parameter infinite degeneracy of vacua labeled by the angle $\theta = \arctan{(y/x)}$. Nonetheless, in this case there is an observable whose value is invariant for all the vacua; this is the radial co-ordinate $r = \sqrt{x^2 + y^2}$; depending on the potential its expectation value is  \unexpanded{$ \langle \hat{r} \rangle = $ constant}. All other observables, such as $\hat{x},\hat{y}, \hat{\theta}$, take on different random values on the different vacua. In our case however, there is no such `invariant observable' apart from the action itself.}. Note that for $\Lambda \leq 0$ and $k \geq 0$, a unique vacuum does exist.

\section{Semiclassical calculations} \label{semiclassical}

Although PIMC  is capable of performing non-perturbative calculations, it is useful to also look at semiclassical theory to see if intuitions are borne out. In this section, we will look at quantum fluctuations around specific  classical solutions for which $A(t)>0$.  The equation of motion in Euclidean time is 
\be
\ddot{A} = -\Lambda A + \frac{2}{3} k A^{-1/3}.
\ee
For $k,\Lambda \neq 0$, this equation cannot be solved analytically so a classical solution must be generated numerically.   For this we restrict attention to   small initial Universes. Numerically this means choosing $A(t_0)=\varepsilon$ where $\varepsilon$ is a small positive number, since the point $A=0$ is singular.

%
\begin{figure}
\includegraphics[width=5in]{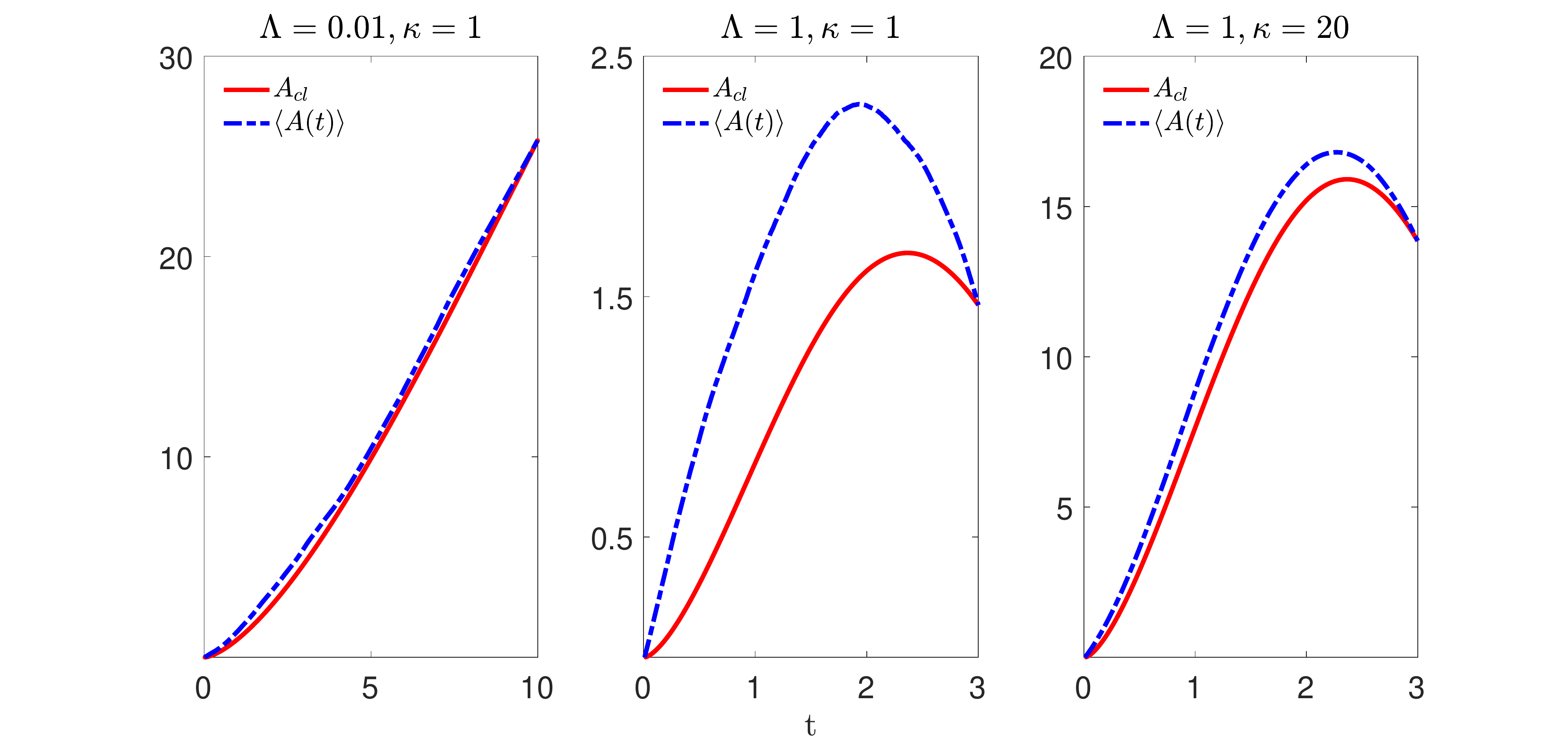}
\caption{The average quantum path $\langle A(t) \rangle$ and the classical solution $A_{cl}(t)$ for various values of $\Lambda$ and $k$. The quantum paths are close to, but distinct from the classical paths.}
\label{semiclassical_paths}
\end{figure}
%

\begin{figure}
\includegraphics[width=5in]{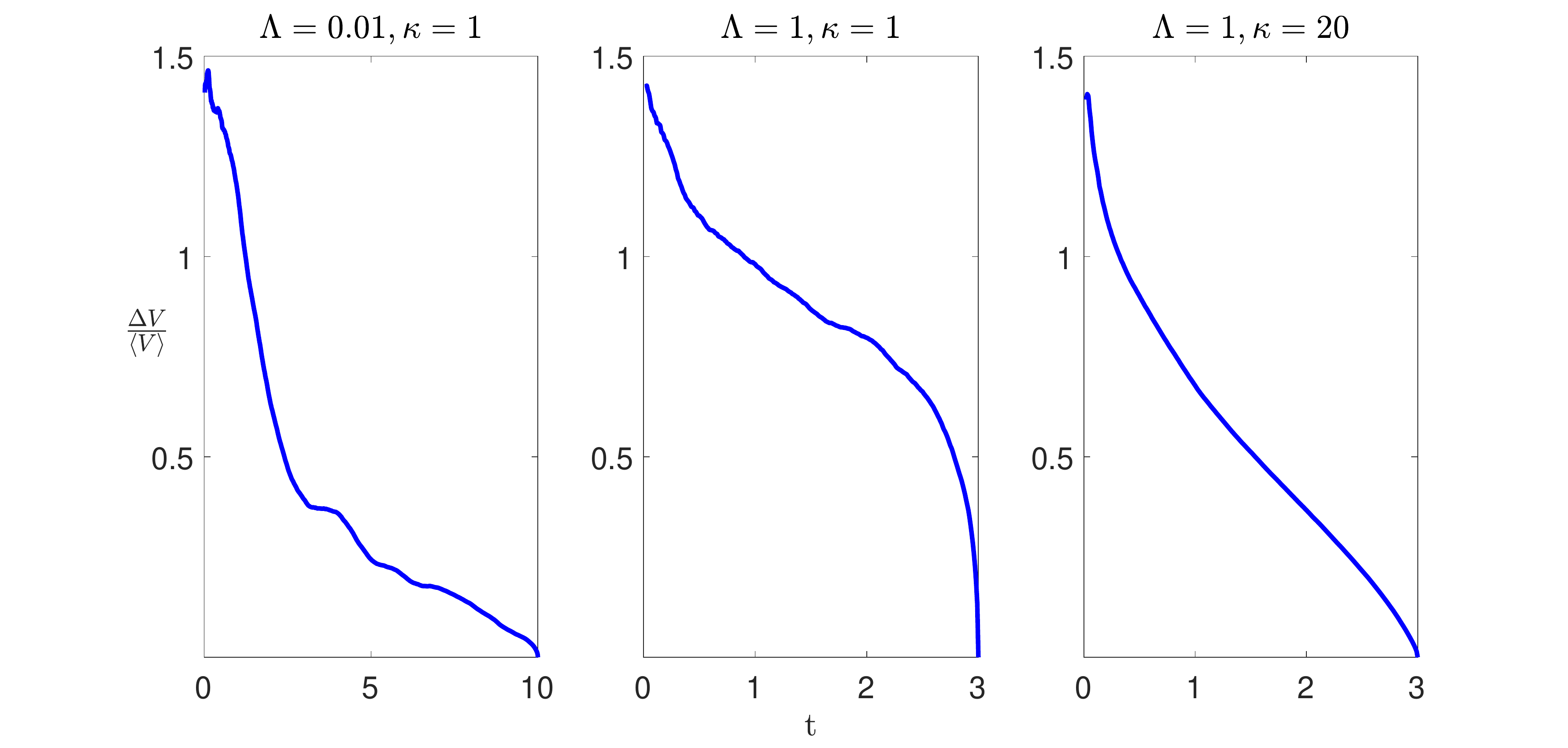}
\caption{Fluctuations in the volume $\Delta V/\langle V \rangle$  for the same paths as in Fig \ref{semiclassical_paths}. It is apparent that at early times and small Universes, fluctuations are large, and gradually reduce as the Universe expands. For $\Lambda=k=1$ (middle), the fluctuations die off slower as compared to others since the classical solution does not expand to sufficiently large volume.}
\label{semiclassical_fluctuations}
\end{figure}

The Monte Carlo process is seeded initially by the classical solution, with the end points $A(0)\approx 0$ and $A(T)>0$  held fixed.  We computed the mean path  $\langle A(t)\rangle$,  and fluctuations in volume $\displaystyle \Delta V(t)/\langle V(t) \rangle\equiv  \sqrt{\langle V(t)^2 \rangle - \langle V(t) \rangle^2}/\langle V(t) \rangle$ for three sets of parameter values $(\Lambda , k) $ . The results are shown in Fig. \ref{semiclassical_paths} and Fig. \ref{semiclassical_fluctuations}. The figures show that (i) the average quantum path does not deviate too far away from the classical path, and (ii) the relative volume fluctuations are large when the Universe is small, and gradually die off as the Universe expands. These results are in accord with expectations, and so provide some verification of our method. Let us however note that the deviation between the  classical and semi-classical path is largest for the case  $\Lambda=\kappa =1$.  This is understood by considering the shape of the potential (\ref{potential}) for various parameter values:  for the first and last frames in  Fig. \ref{semiclassical_paths}, the width of the potential barrier connecting the smaller and larger values of $A$ is significantly larger than for  the center frame. Thus the probability of tunnelling to larger $A$ values, is much larger for the center frame, hence the larger deviation from classicality for $\Lambda=\kappa =1$. 
 
\section{No-Boundary wave function} \label{noboundary}        

The no boundary wave function is given by the path integral  
\be
\psi_{HH}(h, \phi) = \int \D[g]\D[\phi] \exp\l\{-S_E(g,\phi)\r\}
\ee
where the integral is over all compact $4$-geometries bounded by a $3$-geometry with induced metric $h_{ij}$ \cite{Hartle:1983ai}. Since compact geometries are summed over, this integral may be interpreted as the amplitude for the $3$-geometry to arise from ``nothing,'' i.e a zero $3$-geometry or a point. In our  model with dust, the no boundary proposal would correspond to calculating the amplitude of a finite spatial volume $3$-geometry to arise from a zero volume one. That is, we integrate over sets of paths with $A(0) = 0$, with the final value $A(T)\equiv q$ left unspecified; the wave function we calculate is  
\small
\bea
\label{wavefunction}
&&\psi(q,t=T; A_0=0,t=0)  \nn\\
 &&=\int  \D A\, \exp \l\{- \int_0^T dt \l(\frac{\dot{A}^2}{2} - \frac{\Lambda}{2}  A^2 + k A^{2/3} \r) \r \}.
\eea

This proposal is similar to the prescription in \cite{Louko:1988bk} given in canonical coordinates. It may still be considered as a ``no-boundary'' wave function, despite the following differences from the original Hartle-Hawking (HH) proposal: (i) We fix a time gauge and solve the Hamiltonian constraint right at the outset -- there is no integration over the lapse function; we are not seeking solutions of the Wheeler DeWitt equation. (ii) A fixed time gauge also implies that the proper time and the foliation between the initial and final hypersurfaces are fixed. Thus we are integrating over $3-$geometries between these hypersurfaces and not over arbitrary $4-$geometries as in the HH proposal. (iii) As in the Euclidean path integral approach, the integral we consider is also unbounded. However, unlike the HH calculation, the PIMC  gives the full path integral and not a saddle point approximation.  Let us note that  we consider here only the boundary condition $A(0)=0$, and leave  $\dot{A}(0)$ to be determined randomly by the (random) choice of the second value  in the path array, $A(\epsilon)$;  it is of course possible to also fix $\dot{A}(0)$ by fixing the value of $A(\epsilon)$ via the prescription $\dot{A}(0) = (A(\epsilon)-A(0))/\epsilon$, and then randomly selecting the rest of the path. This fixing of two path elements (instead of one) will not make a significant difference to our results for sufficiently long or finely discretized paths.)

As described in the previous section \eq{wavefunction} is discretized as
\be
\label{discretized-wavefunction}
\psi(q,T) = \prod_{i=1}^{N-1}\int dA_i 
\exp \l \{ - \sum_{i = 1}^{N-1} \frac{(A_{i+1} - A_i)^2}{2\epsilon} - \frac{\Lambda}{2}A_i^2 + k A_i^{2/3}   \r\}, \nn \\
\ee
where the sample paths include only those with $A_0 = 0$ and $A_N=q$ left free. The wavefunction $\psi(q,T)$ is determined by binning the values of $q$ at the last time step $T$.  We consider the two distinct cases $\Lambda \leq 0$ and $\Lambda > 0$.

\subsection{$\Lambda \leq 0$}

\begin{figure}
\includegraphics[width = 5in]{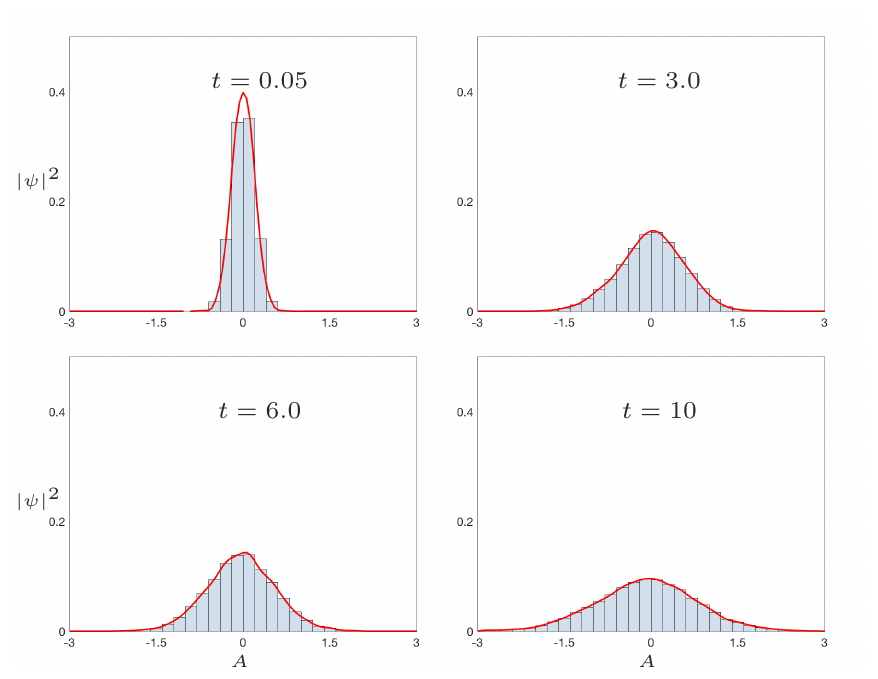}
\caption{No boundary wave function for $\Lambda = -1$, $k=1$: these are snapshots of the ground state wave function at the time slices indicated, for  $T = 10$.}
\label{wavefunction-evolution}
\end{figure}

\begin{figure}
\includegraphics[width = 5in]{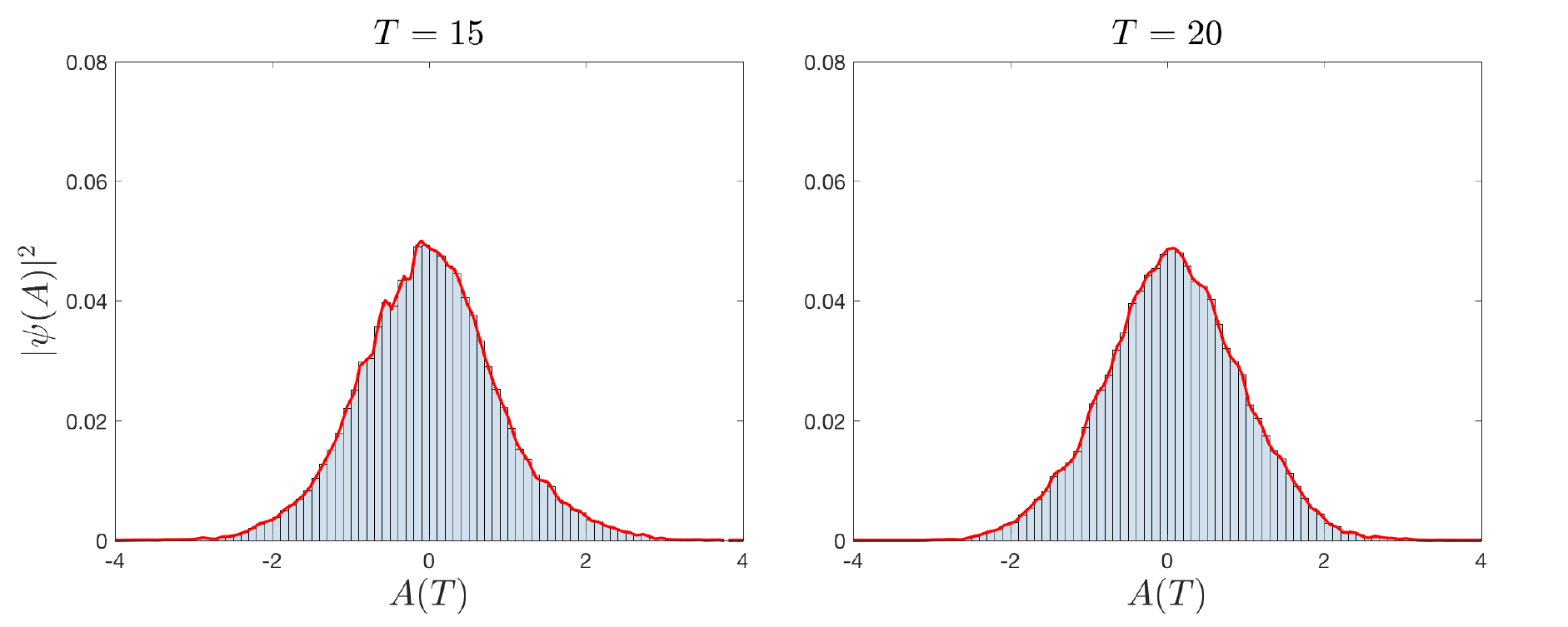}
\caption{The unique ground state wavefunction for the case $\Lambda = -1$. The minor variations in the wave function are due to sample size and bin size. }
\label{unique-wavefunction}
\end{figure}

For this case the Euclidean action  is bounded below so a unique ground state exists. In the large $T$ limit the PIMC algorithm converges to this ground state. The action stabilizes at a finite positive value.  Our results appear in Fig. \ref{wavefunction-evolution} and Fig. \ref{unique-wavefunction}.

%

All plots in Fig. \ref{wavefunction-evolution} are for a fixed final time $T=10$ in the path integral. It displays snapshots of the wave function $\psi(q,t)$ at various $t \le T=10$, obtained by binning paths on the indicated fixed $t$ slices.  It is evident that the wave function starts out highly peaked at $A=0$ for small $t$ and spreads as $t$ increases. The runs from which these plots were made were initiated with a random initial path with $A(t) \in [-500,500]$,  $A(0) = 0$, and parameter values $\epsilon =0.01$, $\Delta = 0.4$ and $ N_{MC} = 50,000$.

%

Fig. \ref{unique-wavefunction} shows the same wave function but at only the final time $T$ in the path integral, $\psi(q,T)$, for $T=15$ and $T=20$. It is evident from this that the results are almost identical, indicating a stable late time wave function.  These plots were again 
 produced from PIMC runs  with a random initial path with $A \in [-500,500]$ and $A(0) = 0$, but this time with  $\epsilon =0.01$, $\Delta = 0.2$, $N_{MC} = 50,000$, and a bin size of $0.1$.

\begin{figure*}
\includegraphics[width = 5.4in]{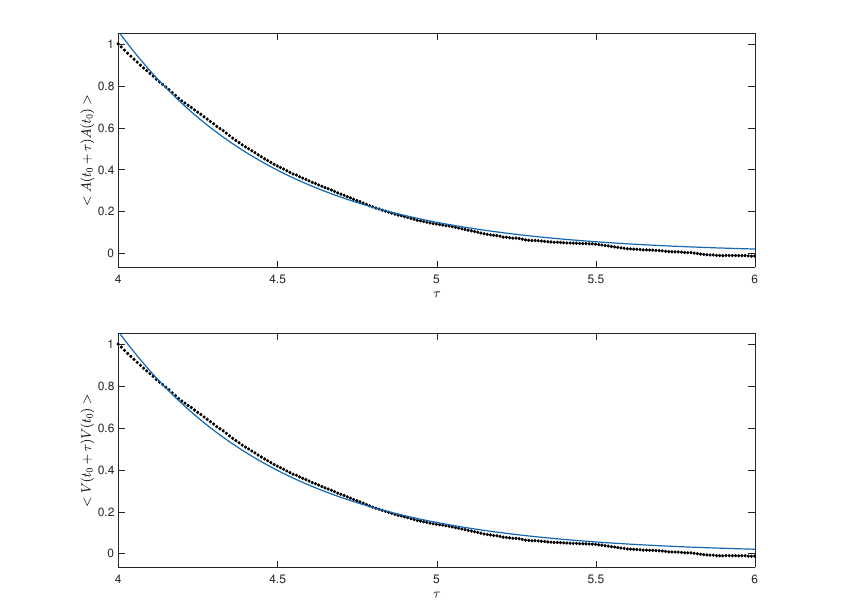}
\caption{Plots of the correlation function for $A(t)$ and the volume $V(t)$ for $\Lambda =-1$. The black dots indicate the actual data points while the solid blue lines indicate exponential curves fitted to the data. Both functions show an exponential decay. }
\label{correlator}
\end{figure*}

%
\begin{figure*}
\includegraphics[width =4.5in]{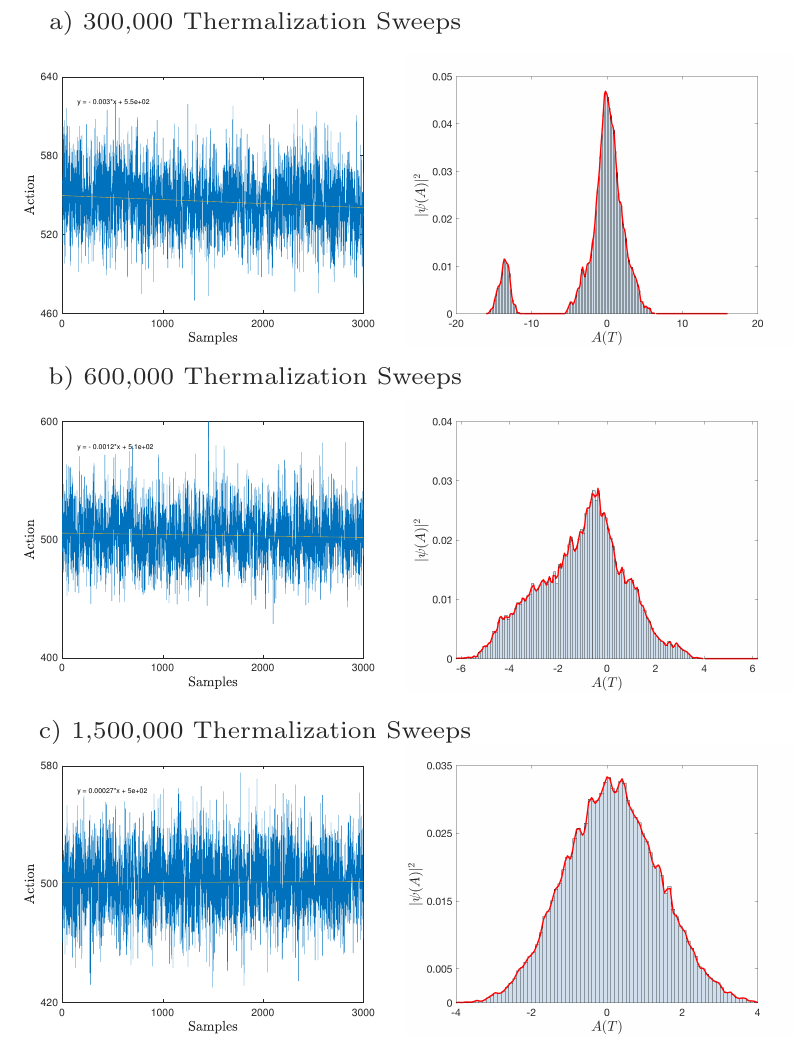}
\caption{The no boundary wavefunction (right panel) for $\Lambda = 0.01$, $k = 1$ and $T=10$ computed using samples collected after varying numbers of thermalization sweeps. A total of $50,000$ samples obtained from $10$ independent MCMC simulations were used for each wavefunction plot. Each simulation was started with a random path with $A(t) \in [-500,500]$ and $A(0) = 0$. The MC parameters were $\epsilon =0.01$, $\Delta = 0.2$ and $N_{MC} = 5000$. The bin size was $0.1$. The left panel shows the value of the action for $5000$ samples from a single simulation. }
\label{thermalization-wavefn}
\end{figure*}

\begin{figure}
\includegraphics[width = 3.5in]{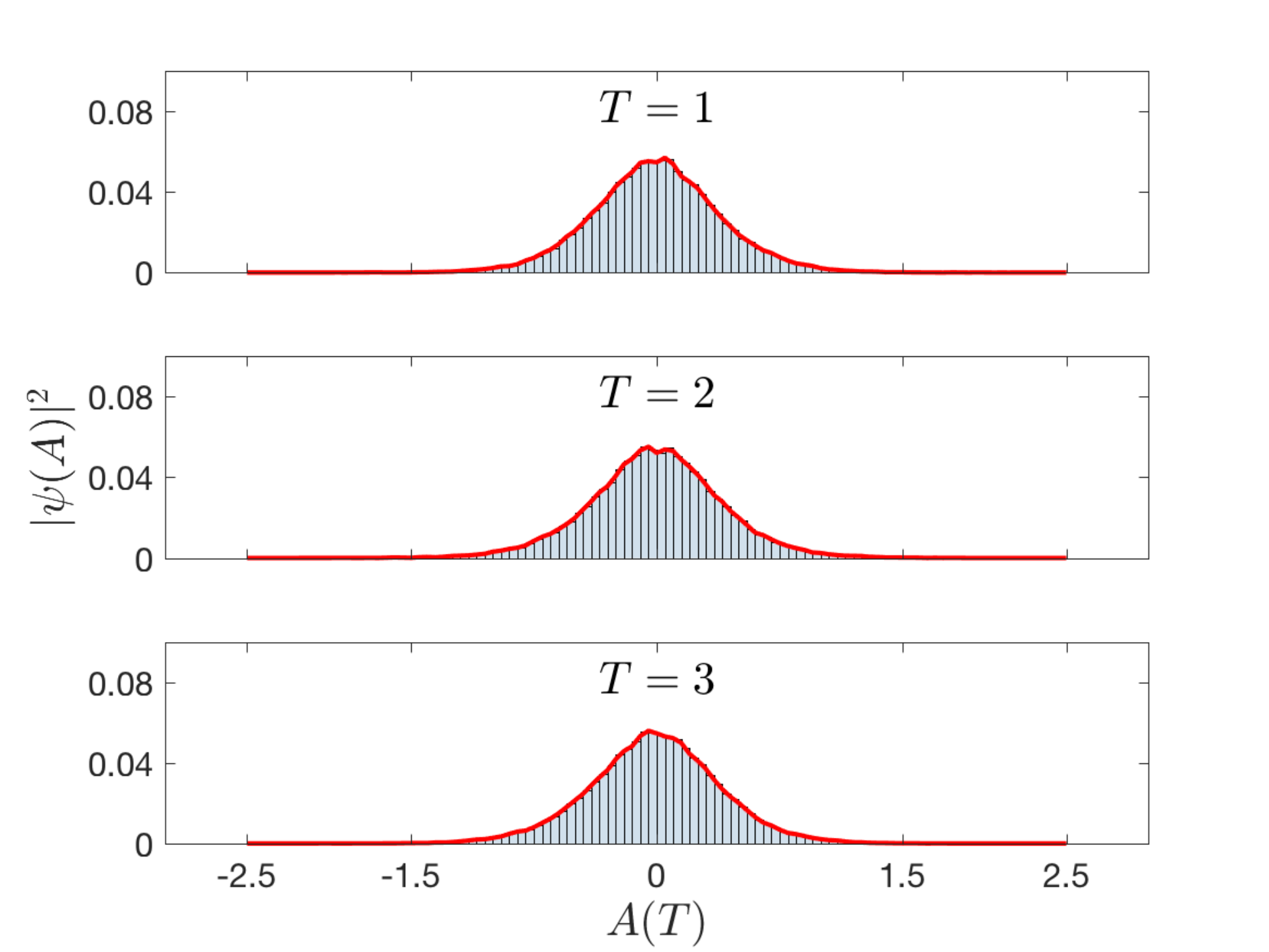}
\caption{The no boundary wavefunction for $\Lambda = 1$, $k = 20$. The plot was generated using $50,000$ sample paths for each value of $T$. The MC runs were started with a random path with $A(t) \in [-500,500]$ and $A(0) = 0$. We used $\epsilon =0.01$, $\Delta = 0.1$ and $ N_{therm} = 10^6$. The histogram was computed using a bin size of $0.05$. }
\label{wavefunctionLone}
\end{figure}

\begin{figure}
\includegraphics[width = 5.5in]{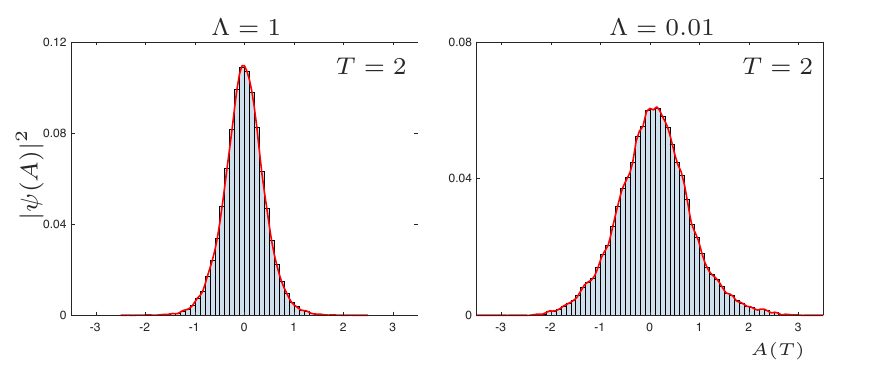}
\caption{The no boundary wavefunctions for $k = 1$, and $\Lambda = 1$ and $\Lambda = 0.01$ for $T=2$. The plot was generated using $50,000$ sample paths. The MC runs were started with a random path with $A(t) \in [-500,500]$ and $A(0) = 0$. We used $\epsilon =0.01$, $\Delta = 0.1 - 0.2$ and $ N_{therm} = 10^6$. The histogram was computed using a bin size of $0.1$. }
\label{varyingLambda_wavefns}
\end{figure}

Lastly for this case, since the ground state is unique, we can calculate the two point correlation function in the state in the usual way by computing
\be
\langle 0| A(t_1)A(t_2)|0\rangle = \frac{ \int \D A\,  A(t_1)A(t_2) \exp\{-S_E\} } { \int \D A \exp\{- S_E \}}.
\ee
This is just the MC average of $A(t_1)A(t_2)$. Fig. \ref{correlator} shows the correlation functions $\langle A(t_0)A(t_0 + \tau)\rangle$ and $\langle V(t_0)V(t_0 + \tau)\rangle$ as functions of $\tau \in [4,6]$ for $T=10$. For comparison, it is worth noting that the volume correlator was also recently computed in the context of the CDT program for $\Lambda >1$ and toroidal spatial slices \cite{Knorr:2018kog}, where the emergent metric is of the FRW form; this work indicates an exponential decay forward in time, similar to what we find here. It is curious that this occurs despite the fact that, unlike in CDT,  we simulate the classically symmetry reduced theory.

\subsection{$\Lambda>0$}
In this case the Euclidean action is not bounded below, but as discussed above, the path integral is convergent for $T < \pi/\sqrt{\Lambda}$. (For the presently observed value of $\Lambda$, this $T$ is in fact very large.)   

For such values of $T$ we can compute  the no-boundary wave function  \eq{wavefunction}. Fig. \ref{thermalization-wavefn} shows the results for $\Lambda = 0.01$, $k=1$ and $T=10$ computed using $50,000$ samples from $10$ independent simulations. The samples were collected after different numbers of thermalization sweeps. The actions for the samples collected after $300,000$ thermalization sweeps and $600,000$ thermalization sweeps are quite similar, both in value and the rate of change, and both appear to be sufficiently thermalized. However, the wave functions computed from these samples are different. After $300,000$ thermalization sweeps, several sample paths appear to take on large values $[A(10) \sim 15, V(10) \sim 84 l_{\text{p}}^{3}]$  whereas after $600,000$ thermalization sweeps the second peak at large $A(T)$ disappears. The tails of the distribution after the $600,000$ thermalization sweeps are significantly thicker compared to the distribution after $1.5$ million sweeps. We tested up to $3$ million thermalization steps and found that  after approximately 1.5 million steps, the slope in the action as a function of time did not change significantly. This is an indication that thermalization in fact did occur. (The cost of our algorithm is order $ N_{\text{therm}} \times N$ where $N$ is the path array size, so the growth in time cost is linear.) Thus, even though the action thermalizes slowly, the trend in Fig. \ref{thermalization-wavefn} suggests that universes of large volume are less probable, at least for these values of $\Lambda$ and $k$. 

It is interesting to compare the last plot in Fig. \ref{thermalization-wavefn} with the plots in Fig. \ref{wavefunctionLone} which shows the no boundary wave function at different values of $T$ with parameters $\Lambda = 1$ and $k = 20$. All three plots in Fig.  \ref{wavefunctionLone} indicate that the universe does not expand to very large volumes within the interval of convergence for the path integral. However, the tails of the wavefunction for $\Lambda = 0.01$, $k=1$ in  Fig. \ref{thermalization-wavefn}  are nearly twice as long as the tails for the wave functions in Fig. \ref{wavefunctionLone}. Fig. \ref{varyingLambda_wavefns} compares the wavefunctions for $T=2$ for both cases. This comparison suggests that larger universes are more probable for smaller values of $\Lambda$.

\begin{figure}
\includegraphics[width = 2.5in]{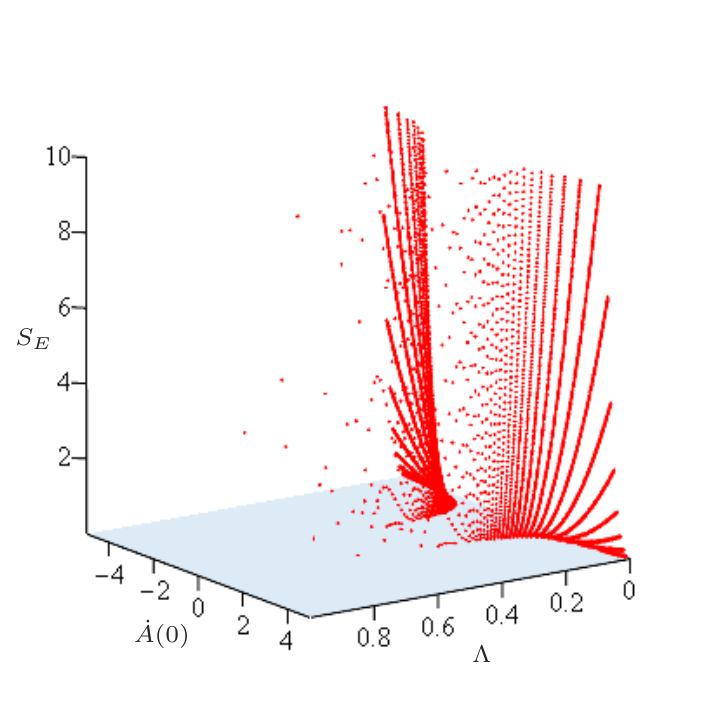}
\caption{A plot of Euclidean action $S_E$  computed on  classical trajectories as a function of $\Lambda$ and initial velocity $\dot{A}(0)$, for $A(0) = 10^{-3}$ and final time $T=10$.  The $S_E = 0$ plane is shaded. Only a small fraction of the $101,100$ classical trajectories displayed here have $S_E =0$. As $\Lambda$ increases, the action for most classical trajectories is negative. This indicates that the vacuum $S_E=0$ surface is almost entirely  populated by non-classical paths.}
\label{classical-vacua}
\vskip 1cm 
\includegraphics[width = 4in]{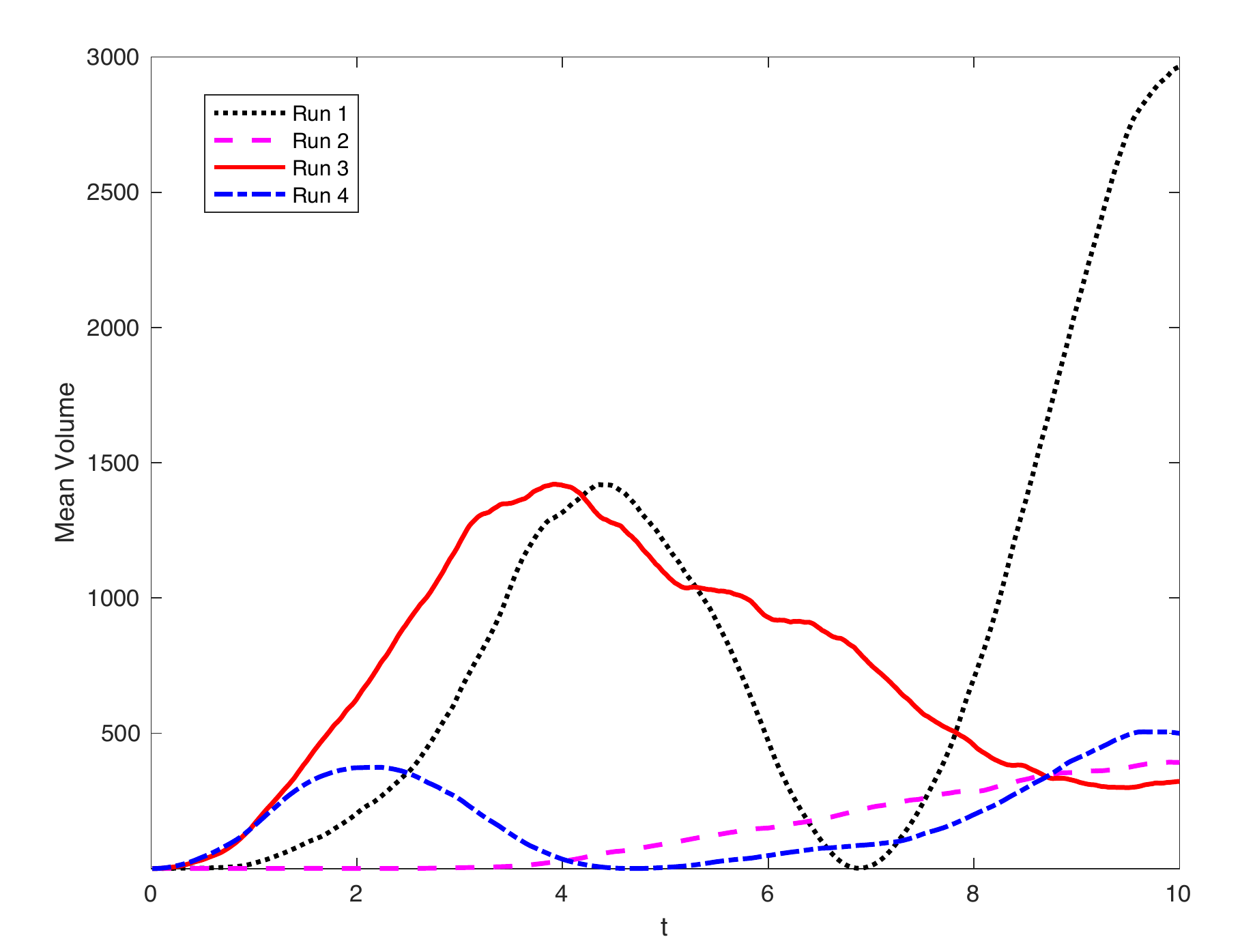} 
\caption{$\langle V(t)\rangle$ for four different MC runs with the same parameters. We used $\Lambda = 1$, $k = 1$ and $\Delta = 0.1$. In each run the algorithm explores a different region of the space of vacua and the results of the runs are never the same. The initial path for each run was a random path with $A(t) \in [-500,500]$ and $A(0) = 0$.}
\label{multirun-vol}
\end{figure}

As discussed above, an  alternatively method for exploring the path integral that is convergent for all $T$ is to use only those paths for which  $S_E \geq 0$.  In general, the set of paths yielding $S_E  = 0$ is uncountably infinite, and the fraction of classical paths in this set is very small. This is displayed in Fig. \ref{classical-vacua}.

In a single MC run only a subset of the $S_E = 0$ paths can be explored. The size and characteristics of this subset are determined by the initial path provided to the algorithm, and the parameter $\Delta$.  As there is no unique ground state, (since numerous physically different configurations have $S_E=0$), the $S_E = 0$ surface defines  a continuum of vacua.  The PIMC algorithm then converges rapidly to the nearest accessible subset of vacua in each run. 

Fig. \ref{multirun-vol} shows the expectation value of volume as a function of time for four different MC runs. Each run was started with a randomly selected initial path and the same value of $\Lambda$, $k$ and $\Delta$. The common feature in the runs is that the volume of the universe is a non-monotonic function of time. In contrast with Figs.  \ref{thermalization-wavefn} and  \ref{wavefunctionLone}, large volumes do arise for initial paths entered starting from zero volume, at least for some period of time. However, each run results in a different trajectory. 

Thus, we see that the two distinct ways of computing the no-boundary wave function we have explored, namely restricting final time $T$ to obtain a convergent result on the one hand, and setting $S_E=0$ on the other, give different physical predictions. The former method is stable and convergent for very small $\Lambda$ (compatible with observations), whereas the latter gives predictions results that are entirely dependent on the initial state due to the large vacuum degeneracy give by the $S_E=0$ bound. 

\section{Tunnelling wave function} \label{tunnel}

\begin{figure}
\includegraphics[width = 2in]{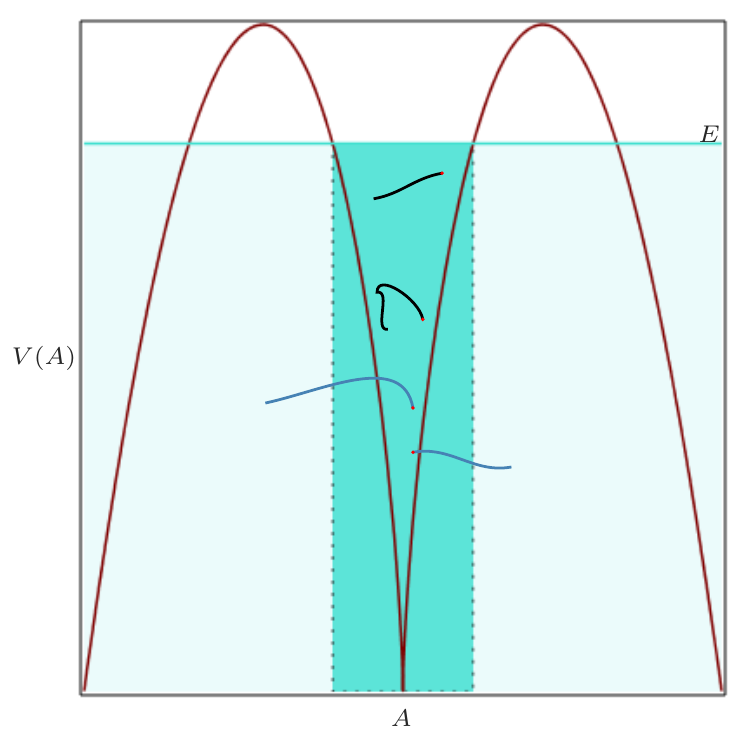}
\caption{This figure illustrates our definition of tunnelling. The paths in black are the classically allowed paths which lie within some energy band $H_P\in[0,E]$. We consider the paths in blue as typical tunnelling paths, since these enter the classically forbidden region. Red dots indicate the starting points of the paths. }
\label{tunnelling}
 \vskip 1cm
\includegraphics[width = 6in]{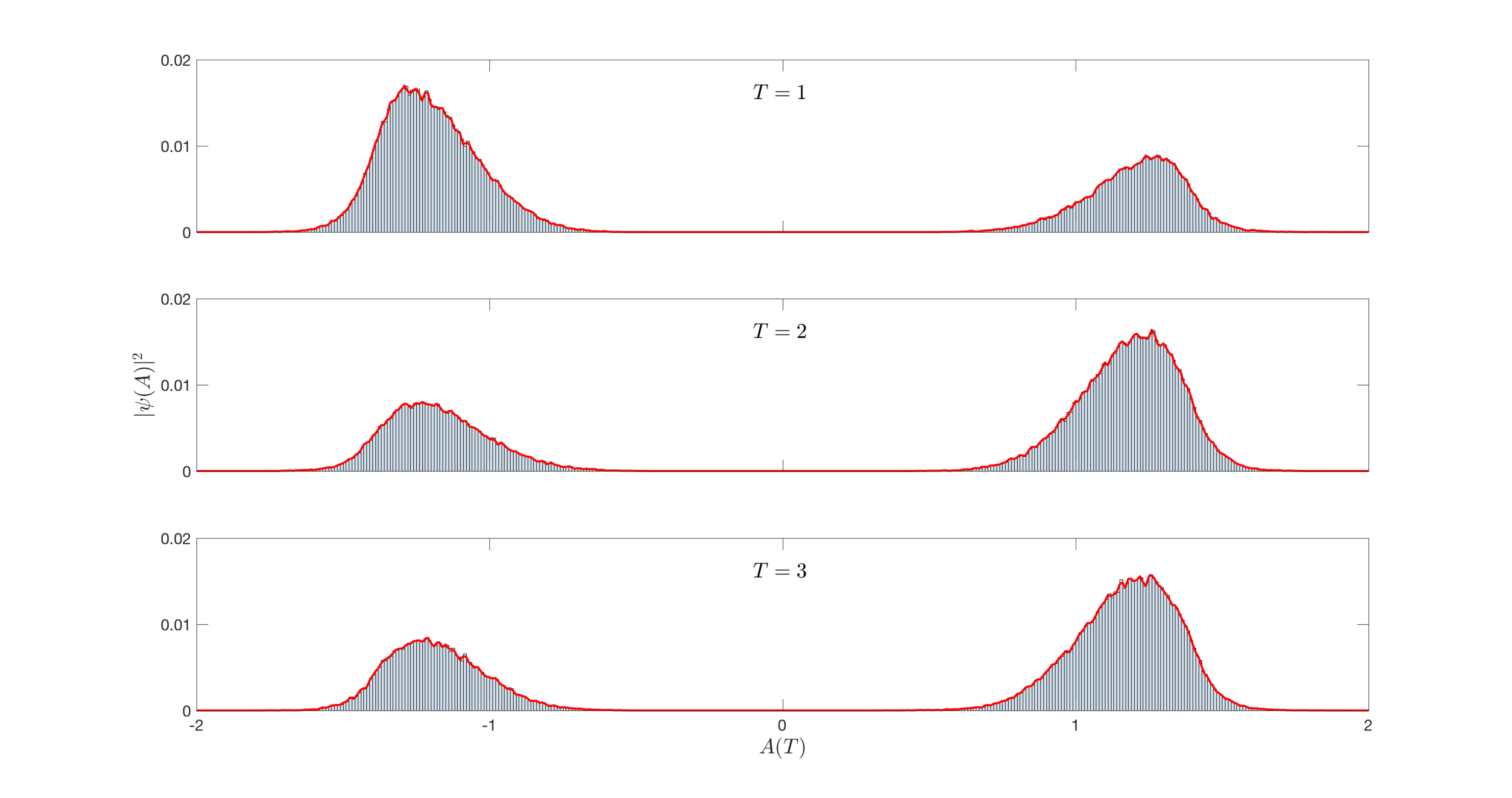}
\caption{Histograms of the final time step for the paths that tunnelled. The tunnelled paths were a subset of $500,000$ samples collected from 10 independent MCMC simulations for each $T$. The model parameters were $\Lambda = 1, k = 20$. The energy band was $[0,24 \text{M}_\text{P}]$. Each run was started with a random path entirely within the classically allowed region. The initial time step ($A(0)$) was sampled but was constrained to remain within the classically allowed region.}
\label{vilenkin-wavefn}
\end{figure}

 The PIMC method we are studying may also be used to  calculate tunelling wave functions  \cite{Vilenkin:1986cy, Vilenkin:1987kf}. The proposal is basically that  the wave function of the universe  include only outgoing waves at the ``boundaries" of superspace. A concrete formulation for this proposal for a deSitter model with a scalar field appears in \cite{Vilenkin:1994rn}. 

We propose here a slightly different notion of tunnelling tailored to our calculation method, and in line with the conventional definition of tunnelling in quantum mechanical systems. This is illustrated in Fig. \ref{tunnelling}. The potential (\ref{potential}) with $\Lambda, k >0$ has two maxima ($V_{max}$). We wish to calculate the wave function for paths with some fixed total energy $H_P=E < V_{max}$ that start with some small value of $A$ within the classically allowed region, and tunnel into the potential barrier beyond the classical turning points within some time $T$.  Since the rejection rates of the MC simulations for paths with fixed energy are quite high, we consider paths in some energy band $H_P\in [0,E]$. These paths are colored blue in Fig. \ref{tunnelling} and are the analogue of the outgoing modes in Vilenkin's proposal. 

The algorithm for the simulation is the same as described in Section \ref{MC-method} except that all paths with pointwise energy outside the band $[0,E]$ are rejected. Another difference in these simulations is that a symmetric discretization is used for the potential as compared to the pointwise discretization in \eq{discrete-action}. The tunnelling region is defined as the region that lies beyond the classical turning points corresponding to the maximum energy $E$ in the energy band. This is represented by the pale blue region in Figure \ref{tunnelling}. 

Our main results appear in Fig. \ref{vilenkin-wavefn}. This displays plots of the wavefunction with parameters $\Lambda = 1, k = 20$ for various values of $T$. The wave function in each case is calculated by binning the final time steps for the subset of paths from the samples that cross the classical turning points at least once. Such paths are deemed to have tunnelled. The total number of samples for each value of $T$ was $5 \times 10^5$.

The main result here is that our simulation method gives explicit tunnelling wave functions. These appear to be anti-symmetric, with one or the other side favoured. This depends on the initial seed path from which the samples are generated.  
\section{Summary and discussion}\label{summary}

We explored the application of the PIMC  method  to FLRW quantum cosmology with dust and cosmological constant. Using the dust field as a clock, we solve the Hamiltonian constraint classically, and then proceed with the application of MC algorithms to compute various properties of the model. Our main result is that the MC method with matter time provides a useful alternative for exploring quantum cosmologies, one that can inform analytical semiclassical calculations of wave functions of the Universe.  

Our specific results are as follows. (i) For $\Lambda \le 0$  the PIMC algorithm converges to the unique ground state wave function for the model; this permits various calculations, such as volume fluctuations and correlation functions. (ii)  For $\Lambda >0$  the potential term in the Hamiltonian is not bounded below. This poses a computational challenge. Nevertheless, for the variables we use, the Euclidean path integral is bounded for final time $T \le  \pi/\sqrt{\Lambda}$ \cite{Carreau:1990is}. Within this bound our results indeed converge; we calculated the no-boundary and tunnelling wave functions. Since the observed value of $\Lambda$ is close to zero, it is clear that this method permits such calculation for very large times. 

We also observed that the alternative method for convergence is to impose a lower bound on the Euclidean action $S_E\ge 0$. Here we saw that there is a continuum of degenerate vacua, and the Markov chain terminates in the neighbourhood of one point of this continuum. For this reason this algorithm does not explore all possible configurations. Indeed {\it any} system with an $S_E$ that is bounded below, and has a continuum of ground states would face this issue in an MC simulation. Thus, since each such vacuum is highly quantum with distinct physical properties, physical results end up depending on the initial state. In our simulations, we see this dependence on initial state appear, for example, in the calculation of the expectation value of the physical volume as a function of time Fig. \ref{multirun-vol}. 

Our approach is readily adapted to include other matter in addition to the dust field. It can also be extended to anisotropic cosmology, where it may prove useful in exploring what becomes of the oscillatory behaviour of scale factors near the singularity. Extending the algorithm to Bianchi I is quite straightforward. Computational cost for a semiclassical analysis would be quite manageable, with three instead of one scale factor paths randomly chosen sufficiently close to a chosen classical path. However, for going beyond semiclassical paths to arbitrary random paths, preliminary simulations indicate that thermalization takes significantly more time. Nevertheless since MC codes are naturally run with parallelization computational cost is not an issue. This work is in progress.    

\begin{acknowledgments}
This work was supported by the Natural Science and Engineering Research Council of Canada. S.M.H. was also supported by the Lewis Doctoral Fellowship.
\end{acknowledgments}

\clearpage
\bibliography{MC}

\end{document}